\documentclass[12pt]{iopart}
\newcommand{\gguide}{{\it Use of $\delta N$ formalism}}
\newcommand{\bfs}{}
\usepackage{iopams}  
\usepackage{graphicx}  
\begin{document}

\title[Use of $\delta N$ formalism]{Use of $\delta N$ formalism\\
- Difficulties in generating large local-type non-Gaussianity during inflation -}

\author{Takahiro Tanaka}
\address{Yukawa Institute for Theoretical Physics, Kyoto University, Kyoto
606-8502, Japan}
\ead{tanaka@yukawa.kyoto-u.ac.jp}
\author{Teruaki Suyama}
\address{
Theoretical and Mathematical Physics Group, Centre for Particle Physics
and Phenomenology, 
Louvain University, 2 Chemin du Cyclotron, 1348 Louvain-la-Neuve,
Belgium}
\ead{teruaki.suyama@uclouvain.be}
\author{Shuichiro Yokoyama}
\address{Department of Physics and Astrophysics, Nagoya University, Nagoya 464-8602,
Japan}
\ead{shu@a.phys.nagoya-u.ac.jp}
\begin{abstract}
We discuss generation of non-Gaussianity in density perturbation through
the super-horizon evolution during inflation by using the so-called
 $\delta N$ formalism. 
{\bfs We first provide a general formula for the non-linearity parameter
 generated during inflation. We find that it is proportional to the
 slow-roll parameters, multiplied by the model dependent factors that
 may enhance the non-Gaussianity to the observable ranges.}
Then we discuss three typical examples to illustrate 
how difficult to generate sizable non-Gaussianity through the
super-horizon evolution during inflation. 
{\bfs First example is the double inflation model,
 which shows that temporal violation of slow roll conditions is not 
enough for the generation of non-Gaussianity. Second example is 
the ordinary hybrid inflation model, which illustrates the importance 
of taking into account perturbations on small scales. 
Finally, we discuss Kadota-Stewart model. This model gives an example 
in which we have to choose rather unnatural initial conditions 
even if large non-Gaussianity can be generated. }
\end{abstract}

\maketitle

\section{Introduction: Success of inflation}
Inflation solved the various cosmological problems like horizon,
homogeneity, flatness problem, just by assuming that the potential
energy of scalar fields dominates the expansion of the universe. 
Moreover, inflation scenario naturally
explains the origin of the almost scale invariant density perturbation. 
The observations in the past decades verified many of the predictions of
inflation. Almost scale invariant spectrum is now confirmed and now we 
even know that the violation of the exact scale invariance is 4\% level~\cite{Komatsu:2010fb}, 
which is also consistent with the simplest slow roll inflation scenario. 

\subsection{Density perturbation}
The amplitude of quantum fluctuation of inflaton $\phi$ 
during slow-roll inflation is determined by the unique relevant mass scale 
at that time, e.g. the Hubble rate $H$:
\begin{equation}
 \delta\phi={H\over 2\pi}.
\end{equation}
However, on scales as large as the Horizon radius 
the meaning of the amplitude of field
perturbation becomes subtle because it depends on the choice of gauge. 
For the flat slicing gauge, in which the trace of the spatial curvature
is kept unperturbed, the perturbation equation for a scalar field becomes
very simple
and looks very similar to the one without gravitational
perturbation~\cite{text}. 
Therefore the amplitude mentioned above can be in fact 
understood as that in the flat slicing gauge. 
This amplitude of perturbation can be interpreted as the dimensionless curvature
perturbation on a uniform energy density surface $\zeta$. Transformation law
is given by 
\begin{equation}
 \zeta= H\delta t=H{\delta\phi\over\dot\phi}={H^2\over 2\pi\dot \phi},
\end{equation}
where $\delta t$ is the shift in time coordinate for this 
transformation, which is given by $\delta\phi/\dot\phi$ 
using the time derivative of the background scalar field. 
Writing the perturbation amplitude in terms of $\zeta$ is useful because 
the curvature perturbation does not evolve on super-horizon scales 
when the evolution path of the universe is unique. 
This constancy of $\zeta$ simplifies the analysis of density
perturbation during inflation for a single inflaton model. 
An extension of this argument to the case of multi-component inflaton is 
the $\delta N$ 
formalism~\cite{Starobinsky:1986fx,Salopek:1990jq,Sasaki:1995aw,Sasaki:1998ug,Lyth:2004gb,Lyth:2005fi}, 
as we explain below. 

\subsection{Further steps from observations}
Further detailed comparison between the theoretical predictions and
future observations is awaited. One important observation will be
tensor-type perturbation in cosmic microwave background. Tensor-type 
perturbation is the transverse-traceless part of the spatial metric
perturbation, which is generated by the same mechanism 
as the perturbation of the inflaton field. Gravitational action 
has a overall prefactor $m_{pl}^2$, and hence we define the canonically 
normalised metric perturbation $\psi_{\mu\nu}=m_{pl}\delta
g_{\mu\nu}$ so as to absorb this factor from the quadratic action. 
Then, the quadratic action for the transverse traceless 
part of $\psi_{\mu\nu}$ becomes identical to the one for a massless 
scalar field. Therefore the amplitude of perturbation $\psi_{\mu\nu}$
generated through almost de Sitter inflation is also $O(H)$. 
Using this relation, we have 
\begin{equation}
 \left\vert{\delta T\over T}\right\vert_{tensor}=O(\delta g_{\mu\nu})
  =O\left({\psi_{\mu\nu}\over m_{pl}}\right)= O\left({H\over m_{pl}}\right).
\end{equation}
Thus we find that the CMB temperature fluctuation caused by 
the tensor perturbation directly probes the value of $H$ during
inflation, i.e. the energy scale of inflation. This can be discriminated 
from the scalar-type perturbation by looking at B-mode 
polarisation~\cite{Mandolesi:2010yw,Baumann:2008aq,quiet}.

Another important observation is the non-Gaussianity of the 
temperature fluctuations~\cite{Komatsu:2001rj,Bartolo:2004if}. 
The non-Gaussianity is caused by 
the non-linear dynamics of cosmological perturbation. 
Once we have completely understood the evolution of density perturbation 
at late time, the remaining non-Gaussianity which is not accounted
for should have their origin in the earlier universe. 
The results of WMAP 7 years indicate that the local-type non-Gaussianity
parameter is given as $f_{NL}=32\pm 21$ at 68\% confidence level~\cite{Komatsu:2010fb}. 
For the Planck satellite, it is expected that 
the window of $f_{{\rm NL}}$ is expected to be reduced to
$O(10)$~\cite{Mandolesi:2010yw}. 

Recently, non-Gaussianity of the primordial
perturbation also has been studied by many authors
~\cite{Lyth:2004gb,Lyth:2005fi,Komatsu:2001rj,
Bartolo:2004if,Maldacena:2002vr,Seery:2005wm,Seery:2006vu,Seery:2005gb,Malik:2003mv,Malik:2005cy,
Yokoyama:2007uu,Yokoyama:2007dw,Byrnes:2006vq,Byrnes:2007tm,Yokoyama:2008by,
Creminelli:2003iq,Alishahiha:2004eh,Koyama:2010xj,Alabidi:2006hg,Malik:2006pm,Sasaki:2006kq,Dvali:2003em,Bernardeau:2002jy,Bernardeau:2002jf,Lyth:2005qk,Salem:2005nd,Seery:2006js,Alabidi:2006wa,ByrnesTasinato,Byrnes:2008wi,Byrnes:2010em,Cogollo,Rodriguez,Enqvist:2004ey,Jokinen:2005by,
Chambers:2007se,Chambers:2008gu,Barnaby:2006cq,Mulryne:2009ci}.
The main reason for non-Gaussianity to attract much attention 
is the expectation to future observations mentioned above. 
These observations may bring us valuable 
information about the dynamics of inflation. 
Here in this short paper, we would like to clarify the difficulties in 
generating large non-Gaussianity from the non-linear dynamics 
during inflation.  

\section{Basic feature of generation of non-Gaussianity of local-type.}
\subsection{Local non-Gaussianity} 

Primordial non-Gaussianity gives rise as non-trivial higher order correlation
functions in primordial perturbations.
In principle, various types of
non-Gaussianity are possibly
generated~\cite{Creminelli:2003iq,Alishahiha:2004eh,Koyama:2010xj}, 
but here we focus on 
the so-called local-type non-Gaussianity. 
Local-type non-Gaussianity is 
characterised by the existence of one-to-one local map between 
the physical curvature perturbation 
$\zeta(x)$ and the 
variable which follows the Gaussian statistics $\zeta_G(x)$ 
at respective spatial points. 
Namely, we can expand the curvature perturbation as 
\begin{equation}
\zeta(x)=\zeta_G(x)+{3\over 5}f_{NL}\zeta_G^2(x)+\cdots,
\end{equation}
and the Gaussian variable $\zeta_G$ satisfies ordinary Gaussian statistics:
\begin{equation}
 \left\langle \zeta^G_{{\bf k}_1}\zeta^G_{{\bf k}_2}\right\rangle
 = \delta^{(3)}({\bf k}_1 +{\bf k}_2 )P_\zeta(k_1). 
\end{equation}

In this case three point function can be characterised by a single
non-linear parameter $f_{NL}$~\cite{Komatsu:2001rj} as  
\begin{equation}
   \left\langle \zeta_{{\bf k}_1}\zeta_{{\bf k}_2}\zeta_{{\bf
  k}_3}\right\rangle
  = \delta^{(3)}({\bf k}_1 +{\bf k}_2 +{\bf k}_3)B_\zeta (k_1,k_2,k_3),
\end{equation}
with 
\begin{eqnarray}
 \!\!\! 
 B_\zeta (k_1,k_2,k_3)={6\over 5}{f_{NL}\over(2\pi)^{3/2}}
 \Biggl[P_\zeta(k_1)P_\zeta(k_2)
+P_\zeta(k_2)P_\zeta(k_3)+P_\zeta(k_3)P_\zeta(k_1)
     \Biggr]. 
\end{eqnarray}
In general, other types of non-Gaussianity can be generated. 
However, when we focus on super-horizon dynamics, 
only local-type non-Gaussianity can be produced as is explained
below. 
\begin{figure}
\begin{center}
\includegraphics[width=80mm]{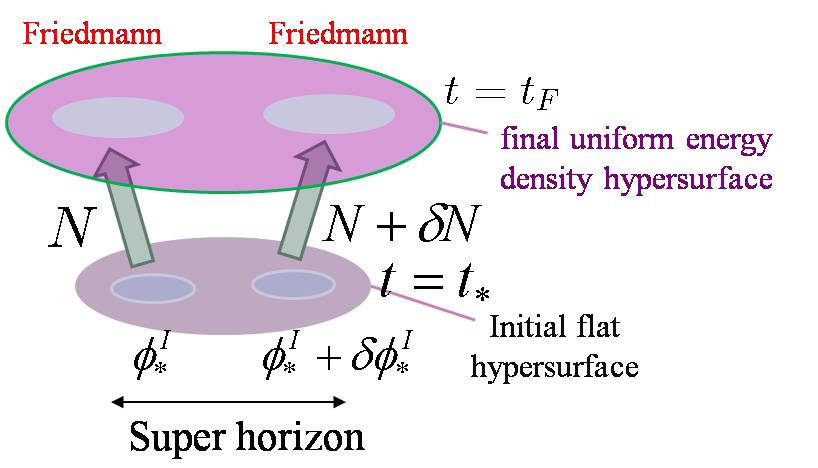}
\caption{Schematic picture to explain $\delta N$ formalism.}
\end{center}
\end{figure}

\subsection{Super-horizon dynamics}  

Here we just present an intuitive explanation about how the density 
perturbations evolve when the length scale is much larger than the 
Hubble scale. Super-horizon dynamics is locally described by the 
Friedmann-Robertson-Walker universe. We consider the evolution of 
the universe starting with an initial 
flat hypersurface at $t=t_*$, on which the initial values of the 
inflaton field have a certain distribution. We evolve the  
spacetime until reaching the final surface at $t=t_F$ 
that is characterised by a specified energy density as shown 
in Fig.~1. Hence, the 
final surface is a uniform energy density surface by definition. 
As each horizon patch is causally disconnected from the others
in the inflating universe, its evolution is determined 
as a local process. 
If initial conditions are completely given in each horizon patch, 
we can solve the evolution of its future. 
Here, for simplicity, we assume that the 
evolution of the averaged values of fields in each horizon patch 
is determined by the averaged values of initial data. This assumption 
is not necessarily true in general, but we have a good reason to 
assume so in most cases. Initial conditions for the smaller scale 
perturbations can affect the evolution of the averaged values of variables. 
However, as there are so many small scale degrees of freedom, 
the average of their effects will not be largely fluctuated 
except for rare situations. 
Of course, even if we can neglect the effect of fluctuations on small scales,  
this does not mean that the backreaction due to small scale 
degrees of freedom does not modify 
the dynamics of the averaged values of variables. This backreaction 
effect arises at the second order perturbation or higher. 
Therefore the difference 
in the backreaction effect between different horizon patches 
is even higher order in perturbation. 
Thus, we usually neglect this effect for 
simplicity. Of course, in the situation like preheating the 
sub-horizon scale perturbations evolve more rapidly than the 
averaged values, we definitely 
need to take the backreaction into account. However, even in 
that case, it will not be necessary to know all the details of 
small scale perturbations to understand the evolution of 
the averaged variables. 

Anyway, we neglect these effects originating from small scale perturbations 
here. Then, the e-folding number between the initial surface and the
final one is completely determined by the initial averaged values 
of the variables. In the scalar field dominant universe, such 
initial conditions are specified by the values of the field components 
and their time derivatives, $\phi^i(t_*,x)$ and $\dot \phi^i(t_*,x)$. 
We denote $\phi^i(t_*,x)$ together with $\dot \phi^i(t_*,x)$ by 
$\phi^I(t_*,x)$. 
Then, the e-folding number between the initial surface
and the final one,  
\begin{equation}
N(t_F,t_*,\phi^I)\equiv \int_{t_*}^{t_F}H dt
\end{equation}
is given as a function of $\phi^I(t_*,x)$. 
Then, roughly speaking, the spatial metric on the final uniform energy
density surface will be given by 
\begin{equation}
ds_{(3)}^2\approx e^{2N(t_F,\phi^I(t_*,x))}
\delta_{ij}dx^i dx^j,
\end{equation}
The curvature perturbation is in fact the perturbation in the above
exponent $N(t_F,t_*,\phi^I(t_*,x))$, and hence we have
\begin{equation}
\zeta(t_F)\approx \delta N(t_F,\phi^I(t_*,x)).
\label{zetadeltaN}
\end{equation}
This formula is the heart of the $\delta N$ 
formalism~\cite{Starobinsky:1986fx,Salopek:1990jq,Sasaki:1995aw,Sasaki:1998ug,Lyth:2004gb,Lyth:2005fi}.

In the slow roll case, the evolution equation for the scalar field 
can be approximated by a first order differential equation in $t$. Therefore 
the phase space of the initial conditions $\phi^I(t_*,x)$ 
is reduced to $\phi^i(t_*,x)$. 
In this case, if there is only a single component, the trajectory 
in the field configuration space, which is one dimensional, is
necessarily unique. 
Then, $\delta N(t_F,\phi^i(t_*,x))$ 
does not depend on the choice of the 
energy density of the final surface. It can be set to a representative 
background value at the initial time. Namely, 
\begin{equation}
 \delta N(t_F,\phi^i(t_*,x))=\delta N(t_*,\phi^I(t_*,x)).
\end{equation}
Therefore nothing non-trivial happens during the super-horizon evolution 
for slow-roll single field inflation.  
Even if the slow-roll conditions are violated, single field inflation 
does not produce sizable $\delta N$ for modes far beyond the horizon
scale. 
This is because the trajectories in the configuration space continues to
converge as the decaying perturbation mode gets irrelevant. 
In order to generate non-Gaussianity through the 
evolution on super horizon scales, 
it is therefore essential to consider multi-component inflaton field. 

Using the formula (\ref{zetadeltaN}), we can express the 
curvature perturbations on the final surface in terms of 
the field perturbations on the initial surface as 
\begin{equation}
\zeta(t_c)= \delta N=
 N^*_I\delta\phi^I_*+{1\over 2}N^*_{IJ}\delta\phi^I_*
 \delta\phi^J_*+\cdots, 
\label{zetaexp}
\end{equation}
where the subscript $*$ indicates the quantities evaluated at 
$\phi^I=\bar\phi^I(t_*)$, and the subscript $I$ associated with $N$  
means the differentiation with respect to $\phi^I(t_*)$:
\begin{equation}
 N_I^*\equiv \left. {\partial N(t_c,\phi^I)\over
	      \partial\phi^I}\right\vert_{\phi=\bar \phi(t_*)}, \qquad
 N_{IJ}^*\equiv \left.
   {\partial^2 N(t_c,\phi^I)\over \partial\phi^I\partial\phi^J}\right\vert_{\phi=\bar \phi(t_*)},  
\end{equation}
where $\bar\phi^I(t)$ is the background trajectory. 
With the aid of Eq.~(\ref{zetaexp}), 
we can write the three point correlation function of $\zeta$ at $t=t_F$
as 
\begin{eqnarray}
\left\langle\zeta({\bf x}_1)\zeta({\bf x}_2)\zeta({\bf
 x}_3)\right\rangle
 & = & N^*_I N^*_J N^*_K 
    \left\langle\delta\phi_*^I({\bf x}_1)\delta\phi_*^J({\bf x}_2)\delta\phi_*^K({\bf
      x}_3)\right\rangle\cr
 && +N^*_I N^*_J N^*_{K L} \Bigl[
    \left\langle\delta\phi_*^I({\bf x}_1)\delta\phi_*^J({\bf
     x}_2)\delta\phi_*^K({\bf x}_3)\delta\phi_*^L({\bf
     x}_3)\right\rangle
\cr &&\qquad\qquad\qquad\qquad    
+(3 {\rm perms})\Bigr]+\cdots. 
\end{eqnarray}
The first term contains only three $\delta\phi$s. If the initial 
fluctuations at $t=t_*$ are Gaussian, this term vanishes. Depending 
on the model of inflation, early generation of non-Gaussianity at
around horizon crossing is possible~\cite{Alishahiha:2004eh}. 
However, the observed almost scale
invariant and slightly red spectrum of initial curvature perturbation 
strongly suggests that the inflation is likely to have been 
in the slow roll regime at 
around the horizon crossing. For the slow roll inflation, it is shown that the early
production of non-Gaussianity is strongly 
suppressed~\cite{Maldacena:2002vr,Seery:2005wm,Seery:2006vu}. 
Therefore, it is well motivated to consider non-Gaussianity contained in 
the second term (or even higher order terms), generated by the
non-linear evolution after the horizon crossing. 

Generation of non-Gaussianity can be classified into two classes. 
One is generation through the super horizon evolution during inflation. 
The other is generation at the end of or after
inflation~\cite{Malik:2006pm,Sasaki:2006kq,Dvali:2003em,Lyth:2005qk,Salem:2005nd,Seery:2006js,Alabidi:2006wa,
Enqvist:2004ey,Jokinen:2005by}. 
In either case the non-Gaussianity is of local type. 
Three point correlation function is characterised by one parameter
$f_{NL}$ alone, and it is related to the e-folding number $N(t_F,\phi^I)$ 
as 
\begin{equation}
{6\over 5}f_{NL}\approx {N^I_* N^J_* N^*_{IJ}\over (N^K_* N^*_{K})^2}, 
\label{fNLformula}
\end{equation}
neglecting higher order terms. 

\section{Non-Gaussianity produced at the end of or after inflation}

There are various models of inflation that produces large
non-Gaussianity of local type at the end of or after inflation.
Here we briefly mention curvaton type
model~\cite{Moroi:2001ct,Lyth:2001nq} as a typical example to explain the mechanism 
to generate large non-Gaussianity. 
Curvaton is another scalar field that is introduced to explain the origin 
of primordial curvature perturbation. During inflation, curvaton field 
is almost massless but it starts to behave as a massive field later when 
the Hubble rate decays down to the mass scale of the curvaton
field. Then, the curvaton starts to oscillate around the minimum of the 
potential and eventually decays into radiation. 
At this point, the fluctuations in the curvaton $\sigma$ are transferred into 
curvature perturbations. For simplicity, we consider the case 
that the density perturbations are dominated by the fluctuations 
originating from the curvaton. 
Neglecting a contribution of second order of 
$\delta\rho_\sigma$ to $\zeta$, which turns out to yield only 
$f_{NL}=O(1)$,
we have~\cite{Sasaki:2006kq}
\begin{equation}
 \zeta = {1\over 4}{\delta\rho_\sigma\over \rho}= 
   {r\over 4}\left(2{\delta\sigma\over\sigma}+\left({\delta\sigma\over\sigma}\right)^2\right),
\end{equation}
where $r=\rho_\sigma/\rho_{tot}$ is the fraction of the energy density 
that the decay products of $\sigma$ takes. Here, we used the estimate 
$\rho_\sigma\approx m^2(\sigma+\delta\sigma)^2/2$ with $m$ being the 
mass of the curvaton. 

The amplitude of the power spectrum is observationally fixed as 
\begin{equation}
P_\zeta=O\left(\left[r{\delta\sigma/\sigma}\right]^2\right)=O( 10^{-10}).
\end{equation}
Therefore, if $\delta\sigma/\sigma$ is not small, $r$ can be as small as
$10^{-5}$.
On the other hand, the non-linear parameter is given by 
\begin{equation}
 f_{NL}=O\left({\langle \zeta^3\rangle/\langle\zeta^2\rangle^2}\right) 
   =O\left({1/ r}\right),
\end{equation}
which can be as large as $10^5$. In this way, the production of 
large non-Gaussianity at the end of inflation is rather easily
achieved. 

{\bfs As a mechanism for getting large non-Gaussianity, 
one can also consider generation of perturbations during preheating
phase~\cite{Tanaka:2003cka,Suyama:2006rk,Enqvist:2004ey,Jokinen:2005by,Chambers:2007se,Chambers:2008gu,Barnaby:2006cq}.
In this case, the background evolution is strongly coupled to the
evolution of small scale fluctuations 
because the e-folding number changes depending on how fast 
the energy of coherent oscillation of fields decays into the small scale 
perturbations. 
While in the other models 
the transmutation 
of homogeneous isocurvature perturbations on each Hubble patch 
into adiabatic perturbations is determined just by solving the background 
evolution of various homogeneous and isotropic universe models. 
In this sense 
the mechanism to generate density perturbation during preheating 
stage is quite different from the other models. 
}
\section{Non-Gaussianity produced during inflation}

Here we just present the formula for $f_{NL}$ in slow-roll inflation 
with canonical kinetic term. 
A systematic derivation for more general formula, including higher 
order correlators, is given in Ref.~\cite{Yokoyama:2008by}. 
Assuming that the initial perturbations of fields 
are Gaussian in the following form 
\begin{equation}
 \left\langle \delta\phi^i_{{\bf k}_1}\delta\phi^j_{{\bf k}_2}\right\rangle
 = \delta^{ij} \delta^{(3)}({\bf k}_1 +{\bf k}_2 )P_{\delta\phi}(k_1), 
\end{equation}
the formula for $f_{NL}$ is written in terms of the potential of the
scalar field $V$ as
\begin{equation}
{6\over 5}f_{NL}=(N^i_* N_i^*)^{-2}
  \int_{N_*}^{N_F} dN'\, N_j(N')Q^j_{kl}(N')\Theta^k(N')\Theta^l(N').
\label{formula}
\end{equation}
Here, we neglected non-Gaussianity in the relation between $\delta N$ and $\delta\phi^i$
at $N=N_F$. 
To define $N_i$ and $\Theta^i$, 
it is convenient to introduce the propagator 
\begin{equation}
\Lambda^i_j(N,N')=\left[T\exp\left(\int_{N'}^N
				P(N'')dN''\right)\right]{}^i_j,
\label{Tproduct}
\end{equation}
where
\begin{equation}
 P^i_j(N)\equiv\left[-{V^i_j\over V}+{V^i V_j\over V^2}
   \right]_{\phi=\bar\phi(N)},
\end{equation}
is the potential term that appears in the linear perturbation equation 
when we use the e-folding number $N$ as a time coordinate. The indices 
associated with $V$ represent differentiation with respect to scalar field. 
$T$ in Eq.~(\ref{Tproduct}) means that the matrices $P^i_j$ 
are ordered in time when the exponential is expanded in power of $P^i_j$. 
Using this propagator, $N_i$ and $\Theta^i$ are defined as 
\begin{eqnarray}
 && N_i(N)=N_j(N_F)
\Lambda^j_i(N_F,N),\cr
&&\Theta^i(N)\equiv\Lambda^i_j(N,N_*)N^j_*. 
\end{eqnarray}
$N_i(N)$ satisfies the equation of motion adjoint to the linear perturbation 
equation for $\delta\phi^i$,
\begin{equation}
{d\over dN}N_i(N)=-P_i^j N_j(N),
\end{equation}
with the boundary condition $N_j(N_F)=(\partial
N(N_F,\phi^i)/\partial\phi^j)|_{\phi=\bar\phi(N_F)}$. 
$\Theta^i(N)$ satisfies the same equation as $\delta\phi^i$ 
\begin{equation}
{d\over dN}\Theta^j(N)=P_i^j \Theta^i(N),
\end{equation}
and the boundary condition is given by 
$\Theta^i(N_*)=N^i(N_*)$, where the index is raised by the inverse of
the field space
metric, which is assumed to be $\delta^{ij}$ here.  
The three point interaction $Q^i_{jk}$ is given by 
\begin{equation}
Q^i_{jk}(N)\equiv\left[-{V^i_{jk}\over V}+{V^i_{j}V_k\over V^2}+
            {V^i_k V_j\over V^2}+{V^i V_{jk}\over V^2}-2{V^i V_j V_k\over V^3}
                    \right]_{\phi=\bar\phi(N)}. 
\end{equation}
In a naive sense, slow-roll conditions require that the
potential of the inflaton is a smooth function of $\phi^i$. 
Hence higher order differentiations with respect to $\phi^i$ 
are suppressed. More precisely, introducing a slow-roll parameter 
\begin{equation}
 \epsilon\equiv {1\over 2}{V_i V^i\over V^2},
\end{equation}
where $m_{pl}$ is set to unity, $P^i_j$ and $Q^i_{jk}$ are supposed 
to be of $O(\epsilon)$ and $O(\epsilon^{3/2})$, respectively.
To obtain an almost scale-invariant spectrum, assuming this kind 
of scaling is natural although it is not strictly required~\cite{Stewart:2001cd}. 
In general, there might be a very massive isocurvature degree of
freedom, which does not satisfy the above simple scaling. 
However, in this simple example, 
fluctuations in such a massive degree of freedom 
decay rapidly, and hence they can be safely neglected. 

Duration of the inflation is roughly estimated as 
\begin{equation}
 N=O\left({V\over dV/dN}\right)=O\left( {H V\over V'\dot\phi}\right)=O\left(\epsilon^{-1}\right).
\end{equation}
Hence, the exponent in the propagator $\Lambda^i_j$ is estimated to be $O(1)$. 
As it is the exponent, it is crucial whether the amplitude of this
factor is slightly bigger 
or smaller than unity. In principle, therefore $\Lambda^i_j$ itself can 
be much larger or much smaller than unity. 
The magnitudes of $N_i$ and $\Theta^j$ depend on the behaviour of the 
propagator, but the value of $N_i$ at $N=N_F$ is given by definition 
so as to satisfy
\begin{equation}
 V(\phi(N_F+\delta N))=V(\bar\phi(N_F)).
\end{equation}
Expanding this equation up to first order, we obtain
\begin{equation}
N_i(N_F)=O\left({V V_i/ V^j V_j}\right) 
=O(\epsilon^{-1/2}). 
\end{equation}
If $\Lambda^i_j$ stays of $O(1)$, one can say that $N_i$ and $\Theta^j$ 
also stay $O(\epsilon^{-1/2})$. Substituting these estimates into 
Eq.~(\ref{formula}), we find that 
$
 f_{NL}=O(\epsilon),
$
as the general argument as given in Refs.~\cite{Maldacena:2002vr,Seery:2005wm,Seery:2006vu} tells. 

However, if amplitude of $\Lambda^i_j$ largely deviates from unity, 
$N_i$ and hence $\Theta^i$ are also largely enhanced or suppressed, 
resulting in large enhancement or suppression of the integrand in 
Eq.~(\ref{formula}). Then, the order of magnitude of $f_{NL}$ is 
not necessarily suppressed like $O(\epsilon)$. Furthermore, slow roll conditions are 
violated, we have more chance to generate large non-Gaussianity.  
However, it is not so easy to construct a viable model which 
generates large non-Gaussianity from the super-horizon dynamics 
during inflation as we shall see below. 

\section{Is there any model which produces large non-Gaussianity 
during inflation?}

\begin{figure}
\begin{center}
\includegraphics[width=60mm]{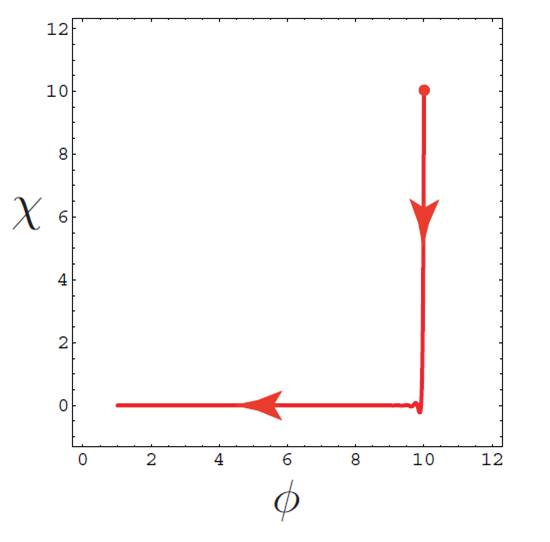}
\label{Fig:trajectory}
\caption{A background trajectory in double inflation
 model. Taken from Ref.~\cite{Yokoyama:2007dw}.}
\end{center}
\end{figure}

\subsection{Temporal slow-roll violation is not sufficient}
\label{doubleinf}
First we consider double inflation model~\cite{Silk:1986vc}, which is a simple two-field
model. The potential is given by 
\begin{equation}
 V(\phi,\chi)={1\over 2}m^2_\phi \phi^2+{1\over 2}m^2_\chi\chi^2.
\end{equation}
There is no direct coupling between two fields, $\phi$ and $\chi$,
except for gravitational one. 

Here we follow the analysis of Ref.~\cite{Yokoyama:2007dw}. 
We show the background trajectory for $m_\chi/m_\phi=20$ 
in Fig.~2 taken from the same reference. 
Since the mass of $\chi$-field is much 
larger, the field moves rapidly in $\chi$-direction first. 
Arriving at around the minimum in $\chi$-direction, the slow roll 
conditions are violated and the field oscillates several times. 
This violation of slow-roll conditions at the intermediate step 
is possible when the mass ratio is large enough. 
Otherwise, the trajectory becomes just a smooth curve and 
slow roll conditions are not violated in the middle. 
One may expect that this temporal violation of slow roll produces 
large non-Gaussianity. 
However, this is not the case. 

During the period when $\chi$-field is moving rapidly, $\phi$-field 
stays almost constant because the mass of $\phi$-field is small 
compared with the Hubble rate which is dominantly determined by the term 
$m^2_\chi\chi^2$ in the potential. Hence, the total e-folding 
number is simply given by the sum of e-foldings for two stages of 
inflation:
\begin{equation}
 N=N^{(\phi)}+N^{(\chi)}\approx \phi_*^2+\chi_*^2.
\end{equation}
Applying the formula (\ref{fNLformula}), we find  
\begin{equation}
{6\over 5}f_{NL}={N_i N_j N^{ij}\over (N_k N^k)^2}={2\over
 \phi_*^2+\chi_*^2}=O\left({1\over N}\right). 
\end{equation}
Hence, as in the usual slow-roll case, the non-Gaussianity is 
suppressed by the slow-roll parameter $\epsilon=O(1/N)$. 

\subsection{Simple hybrid inflation does not give large non-Gaussianity}
\label{hybrid}
As a second example, let's consider hybrid inflation model~\cite{Linde:1993cn}
whose potential is given by 
\begin{equation}
 V(\phi,\chi)={\lambda\over 4}(\chi^2-v^2)^2
   +{m^2\over 2}\phi^2+{g^2\over 2}\phi^2\chi^2.
\end{equation}
Usually $\phi$ is the inflaton field and $\chi$ stays at $\chi=0$ 
during inflation. The effective mass of the $\chi$-field is 
a function of $\phi$ given by 
\begin{equation}
 m^2_\chi=-\lambda v^2+g^2\phi^2, 
\end{equation}
assuming $\chi=0$. The mass of $\chi$-field changes its sign 
at some point, which we denote by $N=N_c$. Then, tachyonic 
instability in $\chi$-direction occurs, leading to the end 
of inflation. 

To analyse the dynamics of $\chi$-field at this critical point 
analytically, we approximate the mass of $\chi$-field as  
\begin{equation}
 m^2_\chi=-\mu^2(N-N_c), 
\end{equation}
where $\mu^2$ is a parameter that depends on the choice of the model 
potential. 

If $\mu^2\gtrsim H^2$, $\chi$-field is massive during $\phi$-field
inflation. Thus, the fluctuation of $\chi$-field decays on large scales. 
If $\mu^2\lesssim H^2$, $\chi$ 
field can stay nearly massless during $\phi$-field
inflation. In this case, the fluctuation of $\chi$-field may play an
important role. As the mass scale of $\chi$ is small, we can apply the 
slow roll approximation to $\chi$-field, too. Then, we have 
\begin{equation}
3H^2\partial_N \chi=\mu^2(N-N_c)\chi.
\end{equation}
This equation can be solved to obtain
\begin{equation}
\ln {\chi\over\chi_*}={\mu^2\over 6H^2}\left[
   (N-N_c)^2-(N_*-N_c)^2 \right].
\end{equation}
Applying the formula $(6/5)f_{NL}\approx N_{\chi\chi}/N_\chi^2$, which is
valid when the fluctuation of $\chi$-field dominates, we have 
\begin{equation}
{6\over 5}f_{NL}\approx {\mu^2\over 3H^2}(N_F-N_c). 
\label{estimatehybrid}
\end{equation}
Since we are restricted to the case with $\mu^2\lesssim H^2$ and 
$N_F-N_c$ cannot be very large, $f_{NL}$ cannot be larger than 
20 or so. If the fluctuations of $\phi$-field also contribute, 
which is usually required in order to produce almost scale invariant 
spectrum, the value of $f_{NL}$ becomes even smaller. 

Furthermore, a more stringent constraint can be set by considering the
following condition. If $\bar\chi(N_c)\gtrsim H$ 
is not satisfied, 
fluctuations at the horizon scale modes at that time determine 
on which side the field rolls down. 
Then, the tachyonic instability does not keep the coherence 
of $\chi$-field on the super horizon scales, 
leading to the so-called spinodal decomposition. 
As a result, long wavelength fluctuation is, roughly speaking, 
completely erased. Hence, there is no chance to generate 
large non-Gaussianity originating from the fluctuation of 
$\chi$-field. The condition $\bar\chi(N_c)\gtrsim H$ becomes 
\begin{equation}
 {\mu^2\over 6H^2}(N_F-N_c)^2\approx \ln {\chi_F\over \chi_c}
   \lesssim \ln {v \over H},  
\end{equation}
where we used the estimate $\chi_F\approx v$.
Using this inequality, the above estimate for $f_{NL}$
(\ref{estimatehybrid}) is bounded by 
\begin{equation} 
{6\over 5}f_{NL}\lesssim {\mu\over H}\sqrt{{2\over 3}\ln{m_{pl}\over H}},
\end{equation}
where we used $H^2=O(\lambda v^4/m^2_{pl})$ and $|m_\chi^2|_{\phi=0}
=O( \lambda v^2)> O(H)$. The latter condition is required to avoid 
topological inflation happens at $\phi=\chi=0$. 
The factor $\sqrt{{2\over 3}\ln{m_pl\over H}}$ is $\approx 7$ even if we reduce 
the energy scale of inflation $\sqrt{m_{pl} H}$ to TeV scale. Here, 
as $\mu^2\ll H^2$ is also required, we conclude that $f_{NL}$ cannot 
be much larger than unity. 

\subsection{Modular inflation}
\label{KS}
\begin{figure}
\begin{center}
\includegraphics[width=70mm]{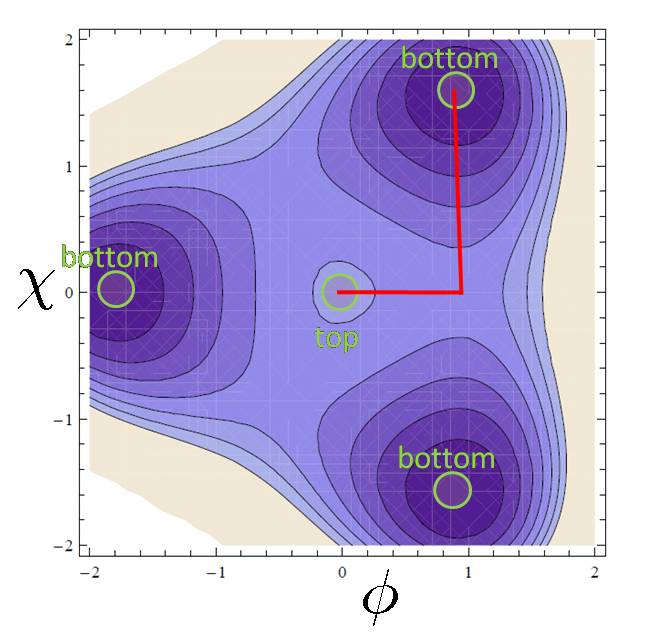}
\caption{Contour plot of the potential of Kadota-Stewart model 
and a typical trajectory.}
\end{center}
\end{figure}

In the preceding subsection we have observed that large mass scale 
for the isocurvature modes 
is necessary to generate large non-Gaussianity. However, the mass 
squared should not be large positive at the early stage of inflation 
when the interesting scales exit the horizon. This requirement 
is not easily compromised unless we invent a very artificial form of 
potential just to produce large non-Gaussianity. 

Here we given one rather natural model which satisfies our requirement. 
This is the model that was proposed by Kadota and
Stewart~\cite{Kadota:2003fs}. 
The basic model potential takes the form 
\begin{equation} 
 V=V_0-m_\phi^2|\Phi|^2
     +{1\over 3}m_\Phi^2(\Phi^3+h.c.)+m_\Phi^2|\Phi|^4,
\end{equation}
where again we set $m_{pl}=1$. $\Phi=(\phi+i\chi)/\sqrt{2}$ is 
a complex scalar field. This potential is modified near $\Phi=0$, the point 
with enhanced symmetry, due to loop correction of the Coleman-Weinberg
type:
$ V_{loop}=-\beta m_\phi^2|\Phi|^2\ln\left|{\Phi/\Phi_*}\right|$.
This correction produces ring-shaped maximum around $\Phi=0$, where 
topological inflation occurs. From the edges of the region of 
topological inflation, the field slowly rolls down the hill. 

Contour plot of the potential is given in Fig.~3. 
We focus on the trajectories which pass near the saddle point 
as shown in the same figure with a red curve. 
In this model mass squared in the $\chi$-direction is negative 
from the beginning. However, since it is the direction of approximate 
$U(1)$ symmetry, the trajectories in any direction are almost equally 
preferred at this stage. As $\phi$ increases, the fluctuation in 
$\chi$-direction 
is enhanced by the ratio of $\phi$ compared with 
its initial value. After the direction of trajectory changes, 
this fluctuation in $\chi$-direction is transferred to the 
curvature perturbation.  
In this way the fluctuations in $\chi$-direction 
can rather easily dominate the final density perturbations. 
{\bfs Then, the amplitude of fluctuations in $\chi$-direction stays 
almost constant near the ring-shaped top of potential, and 
therefore scale invariant spectrum is realised. }

Near the saddle point, the equation in $\chi$-direction can 
be approximated by 
\begin{equation}
3H^2\partial_N\chi=-m_\chi^2 \chi, 
\end{equation}
with $m_\chi^2$ being the mass squared of $\chi$-field evaluated 
at the saddle point. Using the approximation $H^2=$constant,  
this equation can be solved to give 
\begin{equation}
 N={3H^2\over -m_\chi^2}\ln{\chi\over\chi_*}, 
\end{equation}
where $\chi_*$ is the initial value of $\chi$. 
Applying the formula $(6/5)f_{NL}\approx N_{\chi\chi}/N_\chi^2$, 
we obtain
\begin{equation} 
{6\over 5}f_{NL}\approx {m_\chi^2 \over 3H^2}. 
\end{equation}
This can be large if $m_\chi^2 \gg H^2$. 
However, to achieve this by natural form of potential, the 
potential minima as well as the saddles points should be 
located close to the origin in Planck unit. Then, inflation 
does not occur except for the top of the potential 
or near the saddle points. To sustain a sufficiently large 
e-folding number, inflation in these specific regions should last long. 
However, this is not allowed. If inflation lasts long near the 
top of the potential, which is ring-shaped, the motion in the 
radial direction becomes very slow. As a result, the 
amplitude of curvature perturbation originating from the 
fluctuation in $\phi$ direction dominates. For inflation to 
last long near the saddle point, the trajectory should pass 
very close to the saddle point. However, in such cases 
significant fraction of the universe will fall into the 
other side of the saddle point. As a result 
domain walls are formed, which leads to 
the problematic domain wall dominated universe. 

Even if we can avoid the domain wall formation (by considering 
models with higher symmetry), it is not easy to construct 
models which explain the naturalness of such a special trajectory. 
The probability distribution of the universes with different values of 
$\chi_*$ will be affected by the volume expansion factor during 
inflation, which is proportional to $e^{3N}$. If ${3H^2/ |m_\chi^2|}$ 
is small as is requested for large non-Gaussianity, $\chi_*$ must be 
extremely small to gain a large e-folding there.  
Then, the probability of having such a small value of $\chi_*$ is 
extremely small. As long as we consider models which realize 
sufficiently large e-foldings for the natural choice of 
the fiducial value of $\chi_*$, 
$f_{NL}$ will not largely exceed unity. 
Further variation of this model is possible by introducing loop correction 
at around the saddle points, but the final conclusion does not change
much. We will report detailed analysis about this model 
in our future publication.

\section{Conclusion}

Multi-field inflation generates entropy perturbation as well as the
adiabatic one. This entropy perturbation in general affects the
evolution of the curvature perturbation even after the horizon crossing.  
Possible mechanisms of production of non-Gaussianity in the super-horizon 
regime can be classified into two cases. One is the production of
non-Gaussianity during inflation and the other is that at the end of 
or after inflation. Although we could not give a general proof,
it seems very difficult to produce large non-Gaussianity 
during inflation as far as we consider potential without fine-tuning, 
although there are several claims that suggest it
possible
\footnote{
In the examples given in Refs.~\cite{Byrnes:2010em,Cogollo}, 
it is not clear if we
should say that the non-Gaussianity is generated during super-horizon
evolution during inflation 
because distribution in field space stays Gaussian.
In fact, the origin of non-Gaussianity is concentrated in the term
neglected in Eq.~(\ref{formula}).}. 

{\bfs 
Here are several subtle issues. 
In this paper we claimed it difficult to construct a 
natural model in which 
large non-Gaussianity originating from the non-linear evolution 
is produced before inflation terminates. 
By contrast, generation of large non-Gaussianity 
at the end of or after inflation is rather easy. 
However, it may not be so trivial in general to identify when
non-Guassianity is generated. We gave an example of double inflation in 
\S~\ref{doubleinf}. Although we did not mention it in the text, 
in this model the non-linear parameter $f_{NL}$ as a function of 
$t_F$ becomes very large temporally when the background trajectory 
changes its direction, but this large $f_{NL}$ does not persist 
long. This means that the time dependence of $f_{NL}$ is not 
always appropriate to read when non-Gaussianity is mainly generated. 


Another important point to note is the limitation of 
neglecting small scale perturbations. 
The small scale perturbations often become important for the field 
components that become massive during inflation. 
A typical example is the hybrid 
inflation model discussed in \S~\ref{hybrid}. 
To take into account the effect of small scale perturbations, 
$\delta N$ formalism 
based on the background evolution for spatially homogeneous spacetime 
is not sufficient. We need to extend $\delta N$ formalism incorporating
collective variables that characterise statistical state such as 
the averaged magnitude of small scale fluctuation.

Finally we should mention the naturalness of
the initial conditions. It might be possible to obtain large
non-Gaussianity by tuning the initial conditions, but it will not be 
realised in reality if they are extremely fine-tuned. 
As an example, we discussed Kadota-Stewart model in \S~\ref{KS}. 
However, once we start to discuss the likeliness of the chosen 
initial conditions, the long-standing measure problem arises. 
At the moment we are very far from a conclusive answer
on this issue. 
}


\ack
This work is supported by the JSPS through Grants Nos. 19540285, 21244033. 
We also acknowledge the support of the Grant-in-Aid for the Global COE Program ``The Next
Generation of Physics, Spun from Universality and Emergence'' and
the Grant-in-Aid for Scientific Research on Innovative Area N.21111006
from the
Ministry of Education, Culture, Sports, Science and Technology (MEXT) of
Japan.

\end{document}